\documentclass[aps,prl,reprint,groupedaddress]{revtex4-1}

\pdfoutput=1
\usepackage{graphicx}
\usepackage{dcolumn}
\usepackage{bm}
\usepackage{hyperref}
\usepackage{lineno}

\begin{document}


\title{Longitudinal Double-Spin Asymmetries for Forward Di-jet Production \\in Polarized $pp$ Collisions at $\sqrt{s}$ = 200 GeV}


\author{Ting Lin}
\affiliation{Indiana University, for the STAR Collaboration}


\date{\today}

\begin{abstract}
One of the primary goals of the STAR spin program is to determine the contribution of the gluon spin ($\Delta G$) to that of the proton. Recent measurements of the longitudinal double-helicity asymmetry, $A_{LL}$, from inclusive jets place strong constraints on the spin-dependent gluon distribution $\Delta g(x)$ and, for the first time, find evidence for non-zero gluon polarization values for partonic momentum fraction $x$ greater than 0.05. In contrast to inclusive jets, di-jet correlation measurements provide access to partonic kinematics, at leading order, and thus give tighter constraints on the behavior of $\Delta g$ as a function of gluon momentum fraction. Furthermore, di-jet measurements at forward rapidity probe the lower $x$ values where $\Delta g(x)$ is poorly constrained. Preliminary $A_{LL}$ results for di-jets with $-0.8 < \eta_1 < 0.8$ and $0.8 < \eta_2 < 1.8$, from polarized $pp$ collisions at $\sqrt{s}$ = 200 GeV, are presented.
\end{abstract}

\pacs{}

\maketitle

\section{Introduction}
Deep inelastic scattering measurements have found that the spins of the quarks $(\Delta \Sigma)$ account for $\sim30\%$ of the total spin of the proton, and the rest comes from the gluon spin $(\Delta G)$ or orbital angular momentum ($L$) of the partons \cite{RevModPhys.85.655}. Data from the Relativistic Heavy Ion Collider (RHIC) at Brookhaven National Lab have been added to the DSSV \cite{PhysRevLett.113.012001} and NNPDF \cite{Nocera2014276} global analyses. Including the STAR 2009 inclusive jet results \cite{PhysRevLett.115.092002} shows, for the first time, a non-zero gluon polarization in the region of sensitivity, $x > 0.05$.\par
\begin{figure}[h]
  \vspace*{1.0cm}
  \centering
  \includegraphics[width=0.5\textwidth, height=0.5\textwidth]{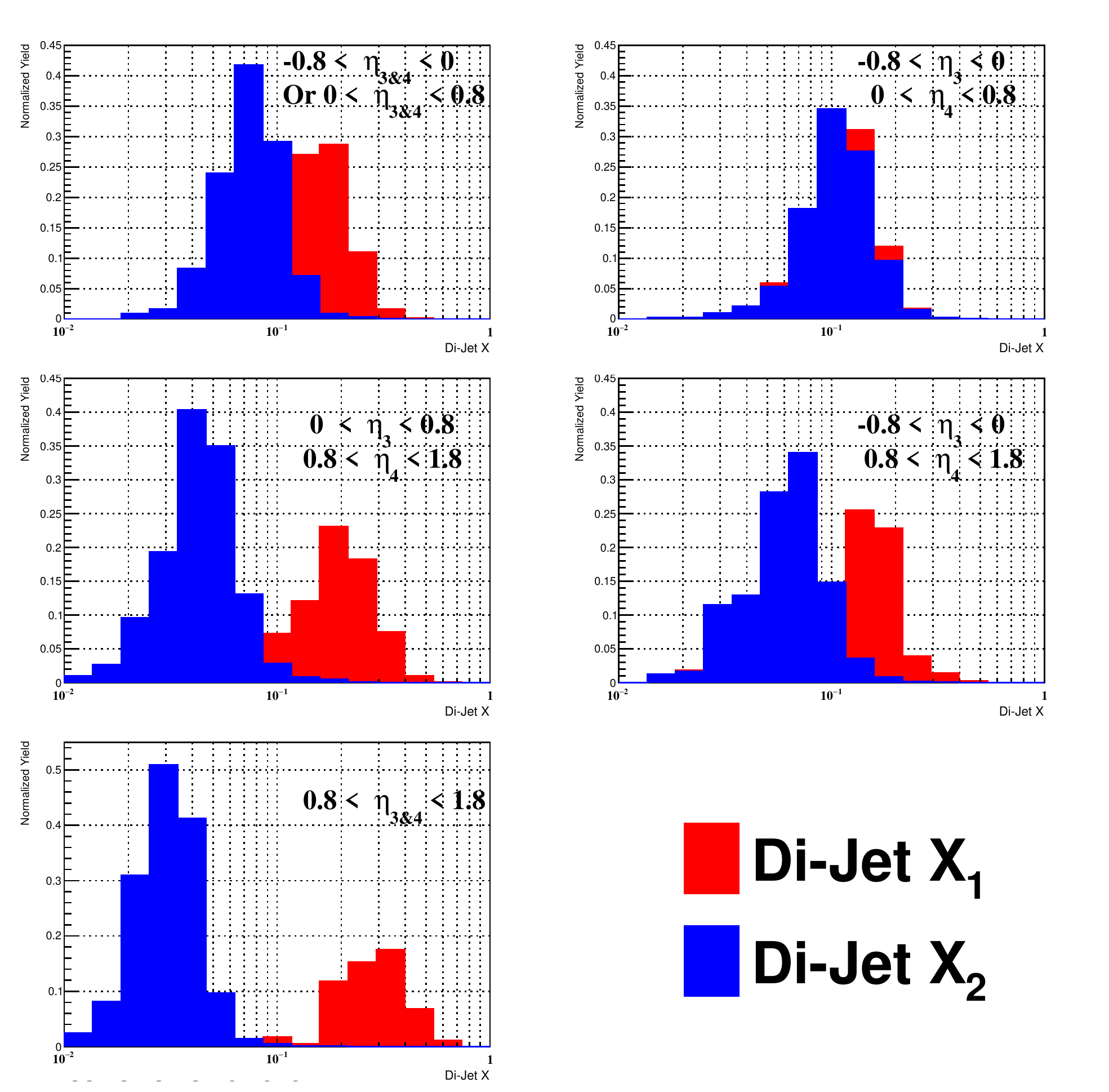}
  \caption{Dijet kinematics: parton $x_1$ and $x_2$ distributions from PYTHIA simulations at $\sqrt{s}$ = 200 GeV for different jet pseudorapidity ranges.}
  \label{fig:Dijet_Kinematics}
\end{figure}

However, the low Bjorken-$x$ behavior and shape of the gluon helicity distribution, $\Delta g(x)$, are still poorly constrained. Correlation observables, such as di-jets, capture more information from the hard scattering and may place better constraints. Di-jet measurements at forward rapidity probe lower $x$ values compared to recent measurements at mid-rapidity \cite{STAR2009pp200Dijets}, as indicated in Fig. \ref{fig:Dijet_Kinematics}.\par
\section{Experiment}
The data used in this analysis were recorded by the Solenoidal Tracker at RHIC (STAR) collaboration in 2009 at $\sqrt{s}$ = 200 GeV with 21 pb$^{-1}$ integrated luminosity. The luminosity-weighted polarizations of the two beams were $P_{B} = 56\%$ and $P_{Y} = 57\%$. The STAR detector subsystems used to reconstruct jets are the Time Projection Chamber (TPC) and the Barrel and Endcap Electromagnetic Calorimeters (BEMC, EEMC) \cite{Ackermann:2002ad}. The TPC provides charged particle tracking in a 0.5 T solenoidal magnetic field over the nominal range $|\eta| \le 1.3$ in pseudorapidity and $2\pi$ in azimuthal angle. The BEMC and EEMC are segmented lead-scintillator sampling calorimeters, which provide full azimuthal coverage for $|\eta| < 1$ and $1.09 < \eta < 2$, respectively. The calorimeters measured electromagnetic energy deposition and provided the primary triggering information via fixed $\Delta \eta \times \Delta \phi = 1\times1$ calorimeter regions (jet patches). A jet patch trigger was satisfied if the transverse energy in a single jet patch exceeded either 5.4 GeV (JP1 trigger, prescaled) or 7.3 GeV (JP2 trigger), or if two jet patches adjacent in azimuth each exceeded 3.5 GeV (AJP trigger). In addition, the Beam-Beam Counters (BBCs) \cite{STARdetector2} were used in the determination of the integrated luminosity and, along with the zero-degree calorimeters (ZDCs) \cite{Ackermann:2002ad}, in the determination of helicity-dependent relative luminosities.
\section{Jet and Di-Jet Selection}
The jet reconstruction procedures follow those used in the 2009 inclusive jet analysis \cite{PhysRevLett.115.092002}. Only tracks with $p_{T} \ge 0.2$ GeV/$c$ and calorimeter towers with $E_{T} \ge 0.2\ \mathrm{GeV}$ were analyzed. Jets were found using the anti-kT algorithm \cite{antikt} implemented in the FastJet \cite{fastjet} package with resolution parameter $R$ = 0.6. Reconstructed jets with $p_{T} > 5$ GeV/$c$ were kept. To avoid double-counting the jet energy contributions from the TPC and calorimeters, towers with tracks pointing to them had the $p_{T}c$ of the track subtracted from the $E_{T}$ of the tower. Negative energies were set to zero.\par
For each event, di-jets were selected by choosing the two jets with the highest $p_{T}$ that fell in the pseudorapidity range $-0.8 \le \eta \le 1.8$. Other conditions were the same as those used in the 2009 di-jet measurements at mid-rapidity \cite{STAR2009pp200Dijets}.
\section{Simulation}
To correct for detector effects on the measured jet quantities, simulated events generated from PYTHIA 6.425 \cite{PYTHIA} with the Perugia 0 tune \cite{Perugiatune} were chosen to run through a STAR detector response package implemented in GEANT 3 \cite{Geant}. The simulated events were embedded into ‘zero-bias’ data, which were triggered on random bunch crossings over the span of the run, allowing the simulation sample to track accurately the same beam background, pile-up, and detector conditions as the real data set.\par
\begin{figure}[h]
  \centering
    \includegraphics[width=0.23\textwidth]{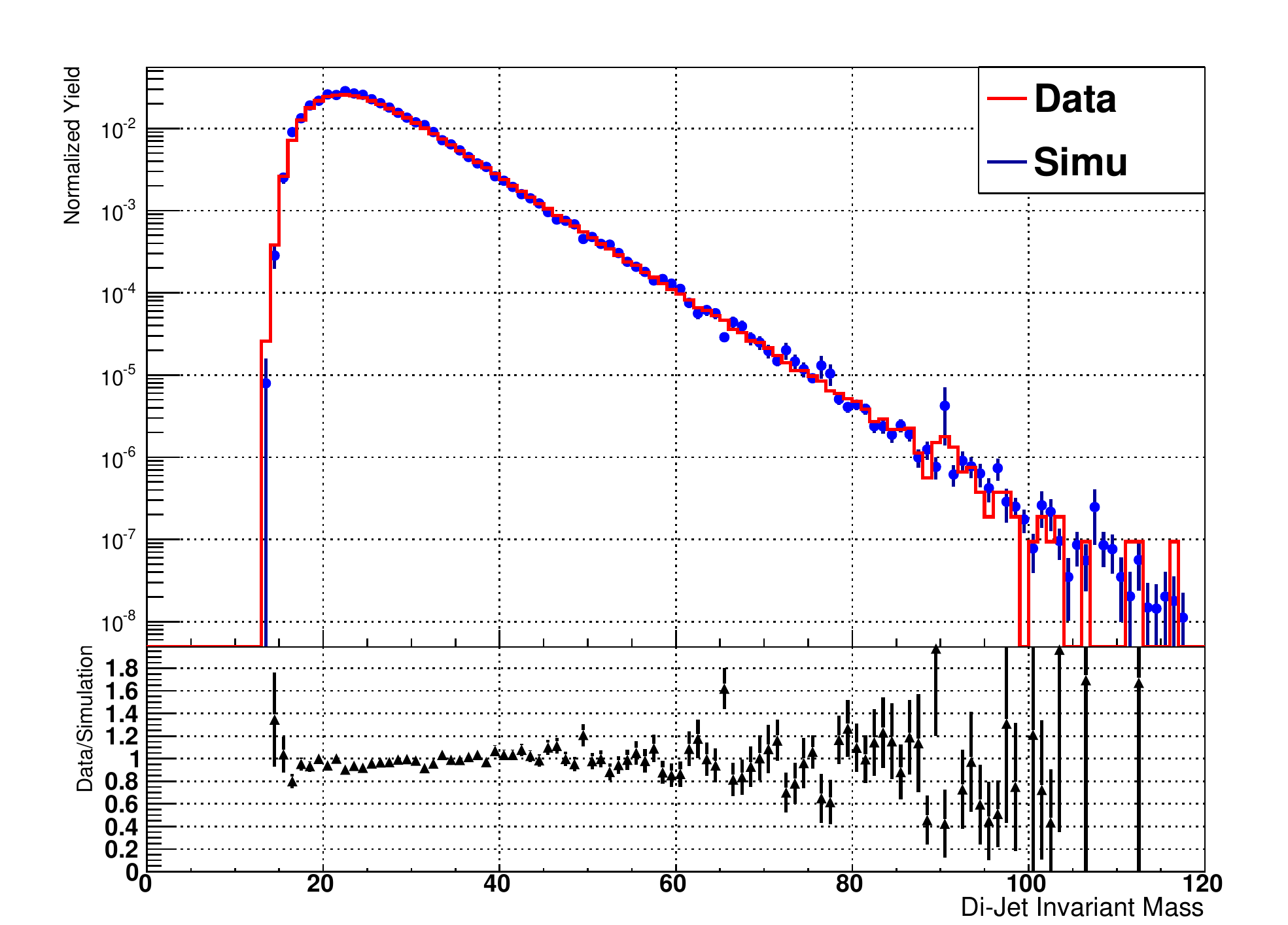}
    \includegraphics[width=0.23\textwidth]{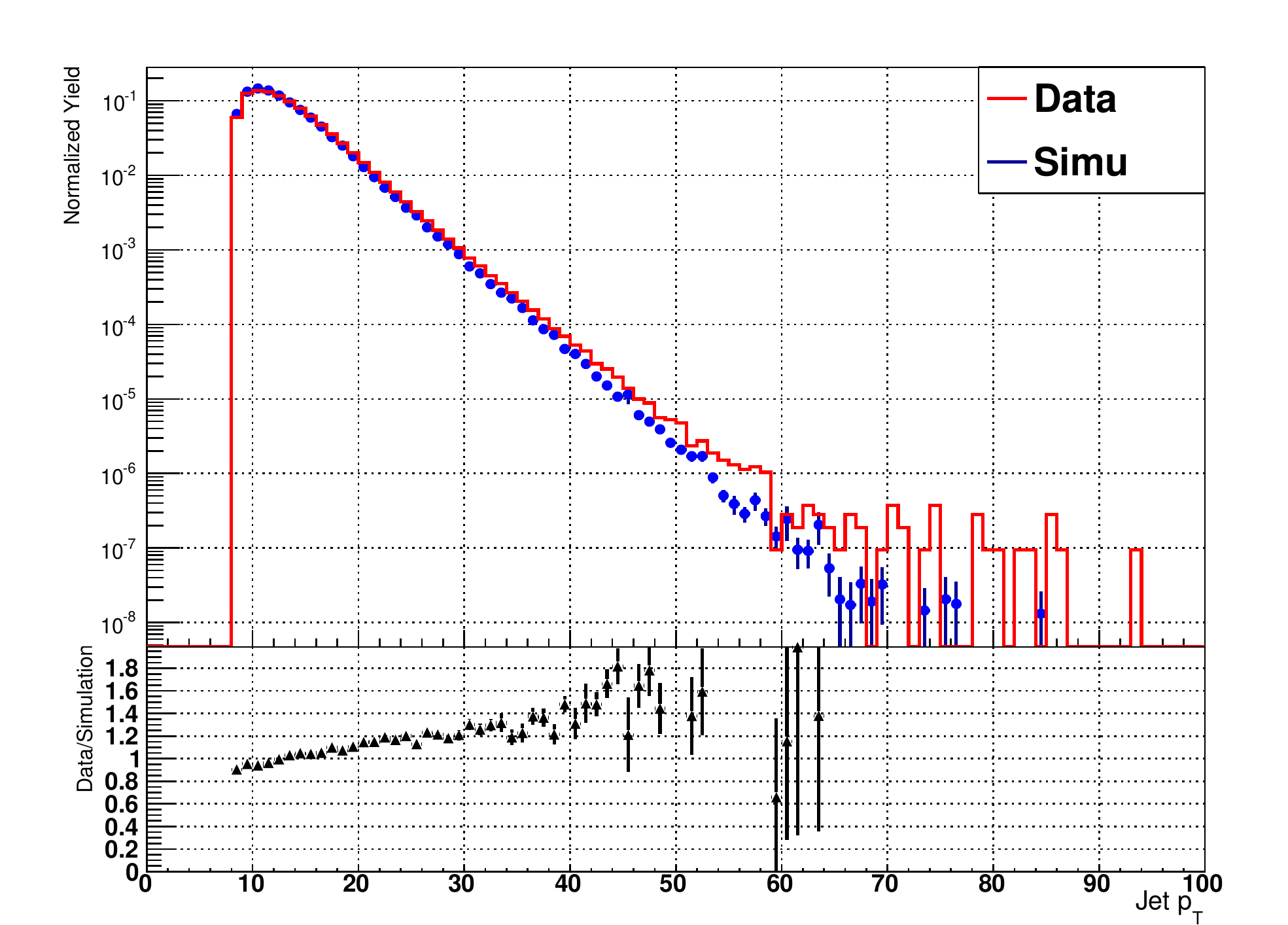}
    \includegraphics[width=0.23\textwidth]{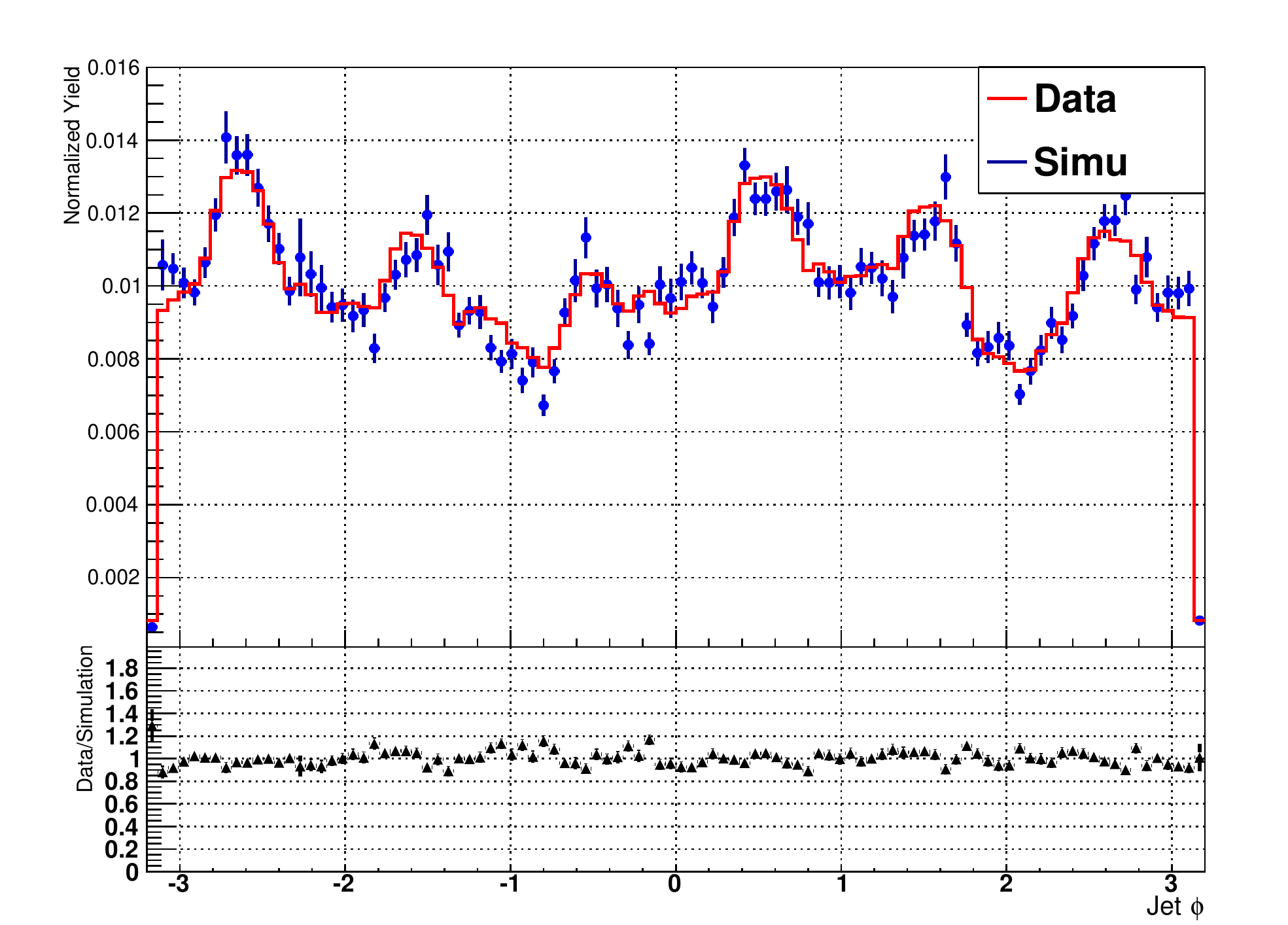}
    \includegraphics[width=0.23\textwidth]{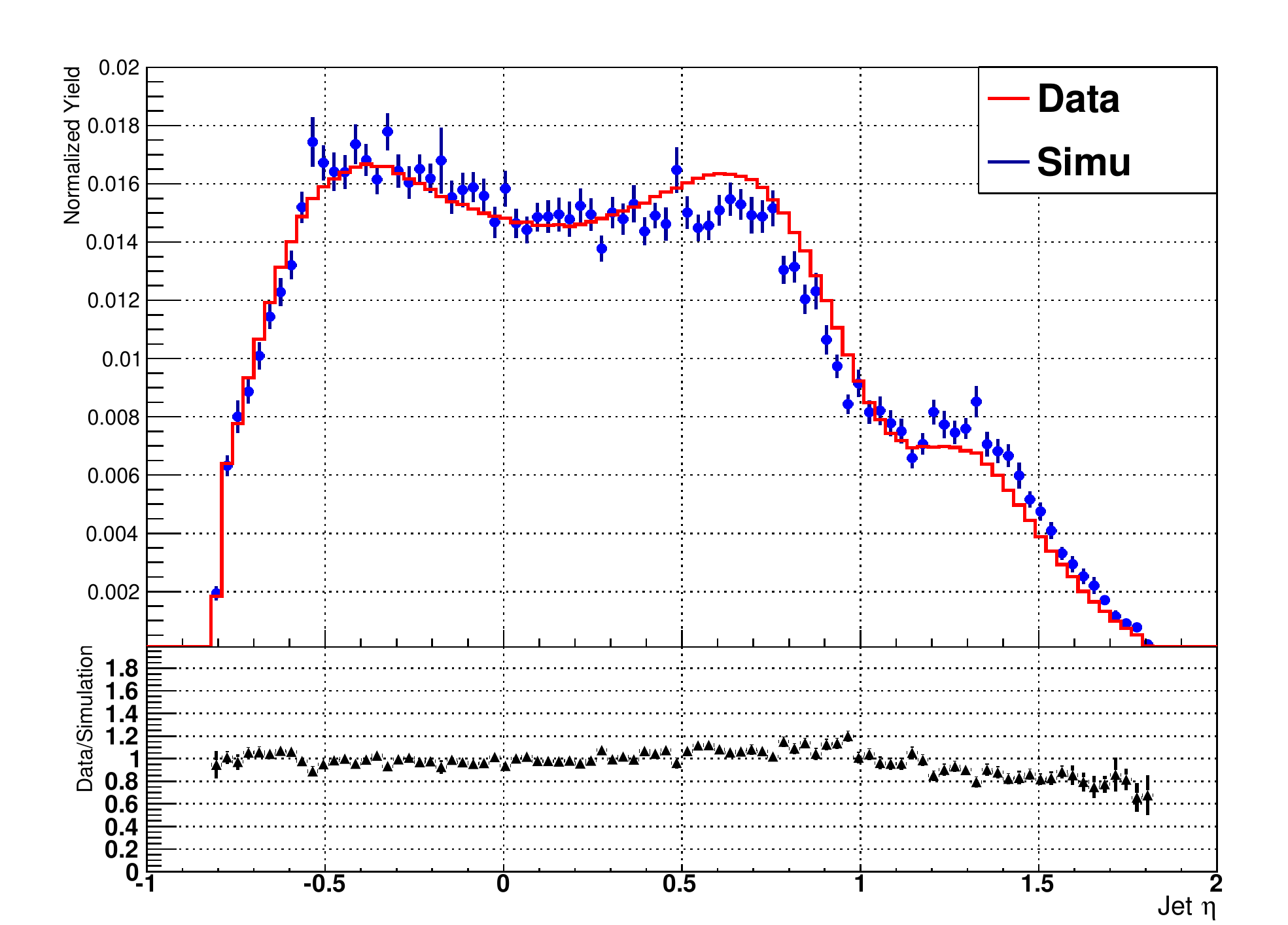}
    \caption{Comparison of jet and di-jet yields as a function of invariant mass (upper left), $p_{T}$ (upper right), azimuthal angle (lower left) and pseudorapidity (lower right) between simulation and data, showing good agreement}
    \label{fig:Dijet_DataSimu}
\end{figure}

Detector-level di-jets were reconstructed using the same algorithms as the data. Comparison of the jets and di-jets reconstructed from data and simulation confirms that the STAR detector response is well understood, as indicated in Fig. \ref{fig:Dijet_DataSimu}. Di-jets were also reconstructed in simulation at the particle level, which were formed from final-state stable particles produced in the simulated event.
\section{Challenges and Methods}
\begin{figure}[!h]
  \centering
    \includegraphics[width=0.45\textwidth]{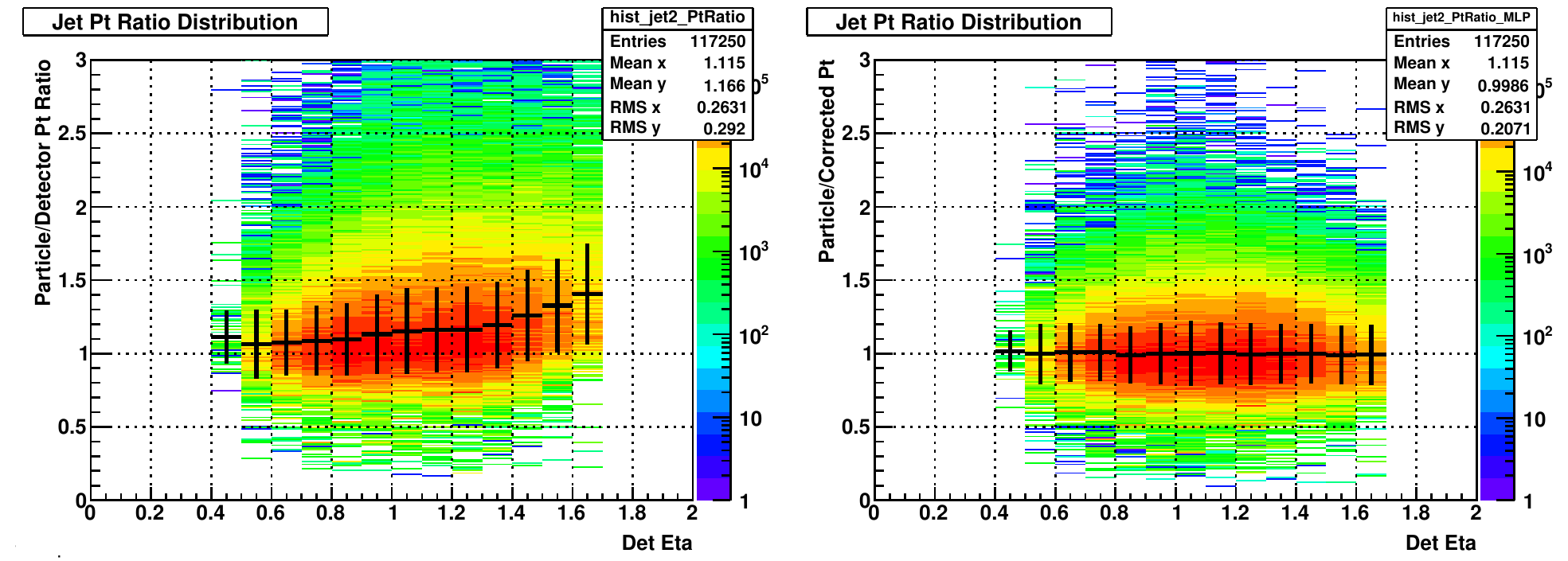}
    \caption{Jet particle-level $p_{T}$ over detector-level $p_{T}$ as a function of detector $\eta$ before (left plot) and after (right plot) a $p_{T}$ shift correction is applied. The correction is determined using machine-learning techniques}
    \label{fig:Jet_ShiftPt}
\end{figure}

As noted before, the STAR TPC only covers the nominal range $|\eta| \le 1.3$; the tracking efficiency decreases rapidly in more forward regions. Lower tracking efficiency means the reconstructed jets will have lower $p_{T}$ on average, as shown in the left panel of Fig. \ref{fig:Jet_ShiftPt}. This inaccurate $p_{T}$ reconstruction skews the extraction of the initial state parton momenta. Jets with a high percentage of neutral energy are preferentially selected, both in triggering and reconstruction, leading to a biased sample.\par
A machine-learning regression method was used to shift the jet $p_{T}$ back to its true value. Such supervised machine-learning regression algorithms make use of training events, for which the desired output is known, to determine an approximation of the underlying functional behavior defining the target value.\par
Multilayer Perceptron (MLP from TMVA \cite{TMVA}) is the main algorithm applied in this analysis; see Fig. \ref{fig:Jet_ShiftPt} right panel. Systematic errors are evaluated and added in quadrature from the output using different training sample sizes, changing the number of layers and nodes, and using the Linear Discriminant algorithm (LD from TMVA) as an alternate method.\par
The di-jet invariant mass is also skewed during the jet reconstruction. In the jet-finder algorithm, tracks are assigned the mass of charged pions, while for the EMC towers the particles are assumed to be massless, which makes the detector-level jet invariant mass lower than its real value. Hence, similar corrections are also made here to shift the jet invariant mass back to its true value (particle level).\par
Barrel and Endcap jets are separately corrected in $p_{T}$ and mass, and di-jet invariant masses are calculated using the shifted jet transverse momentum and mass from machine learning.\par
\section{Results}
Sorting the yields by beam spin state enables a determination of the longitudinal double-spin asymmetry $A_{LL}$, evaluated as 
\begin{equation}
A_{LL} = \frac{\sum(P_{Y}P_{B})(N^{++}-rN^{+-})}{\sum(P_{Y}P_{B})^{2}(N^{++}+rN^{+-})},
\label{equ:ALL}
\end{equation}
where $P_{Y,B}$ are the polarizations of the yellow and blue beams, $N^{++}$ and $N^{+-}$ are the di-jet yields from beam bunches with the same and opposite helicity configurations, respectively, and $r = \mathcal{L}^{++}/\mathcal{L}^{+-}$ is the relative integrated luminosity of these configurations. The sum is over individual data runs, which ranged from 10 to 60 minutes in length and were short compared to changes in beam conditions.\par
\begin{figure*}
  \centering
  \begin{minipage}{.495\textwidth}
    \centering
    \hspace*{-0.6cm}
    \includegraphics[width=1.03\textwidth,height=0.55\textwidth]{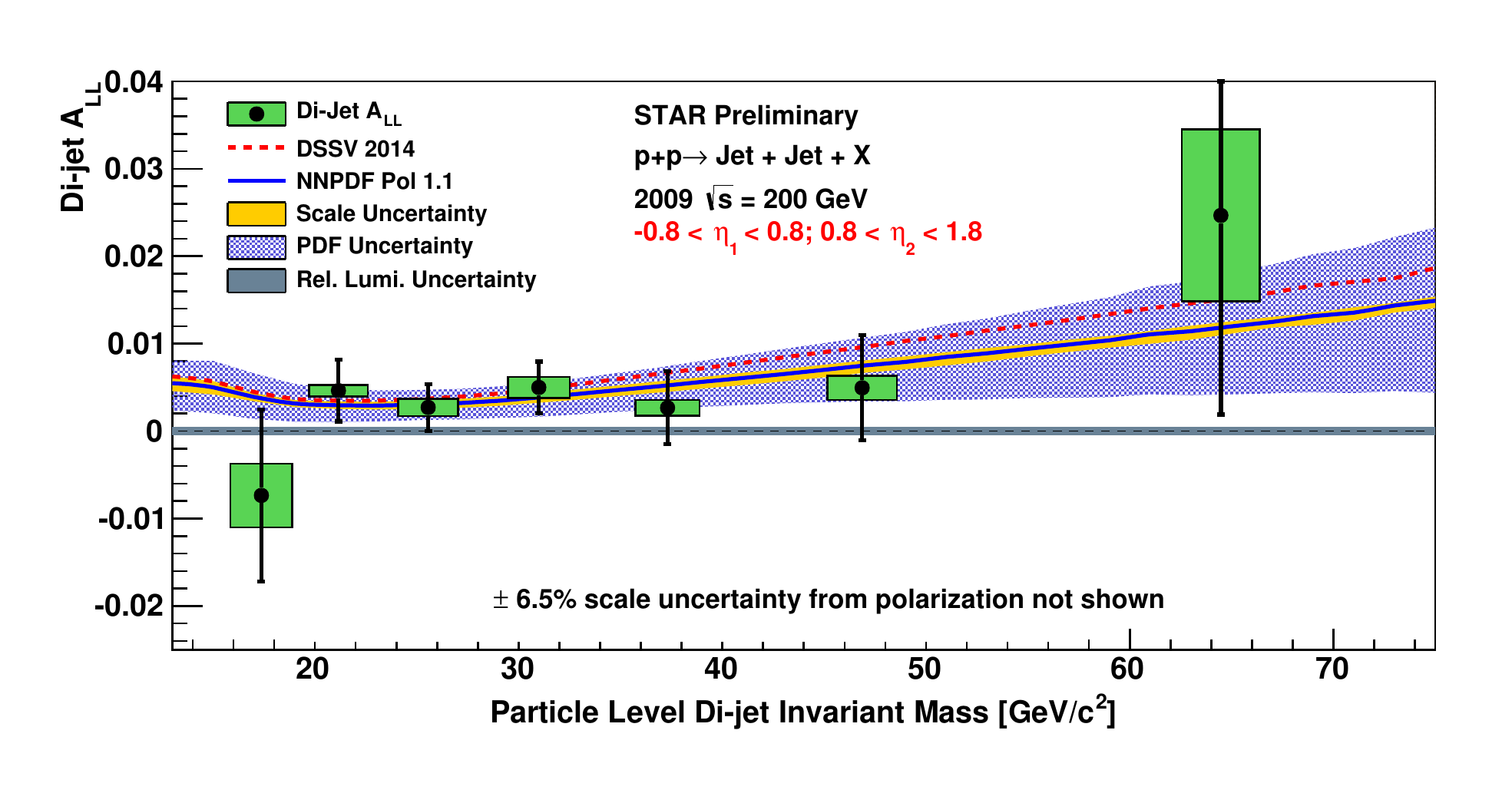}
    \caption{Preliminary Di-jet $A_{LL}$ for Barrel-Endcap full topology}
    \label{fig:ALL_BE_FullTopo}
    \hspace*{-0.6cm}
    \includegraphics[width=1.03\textwidth,height=0.55\textwidth]{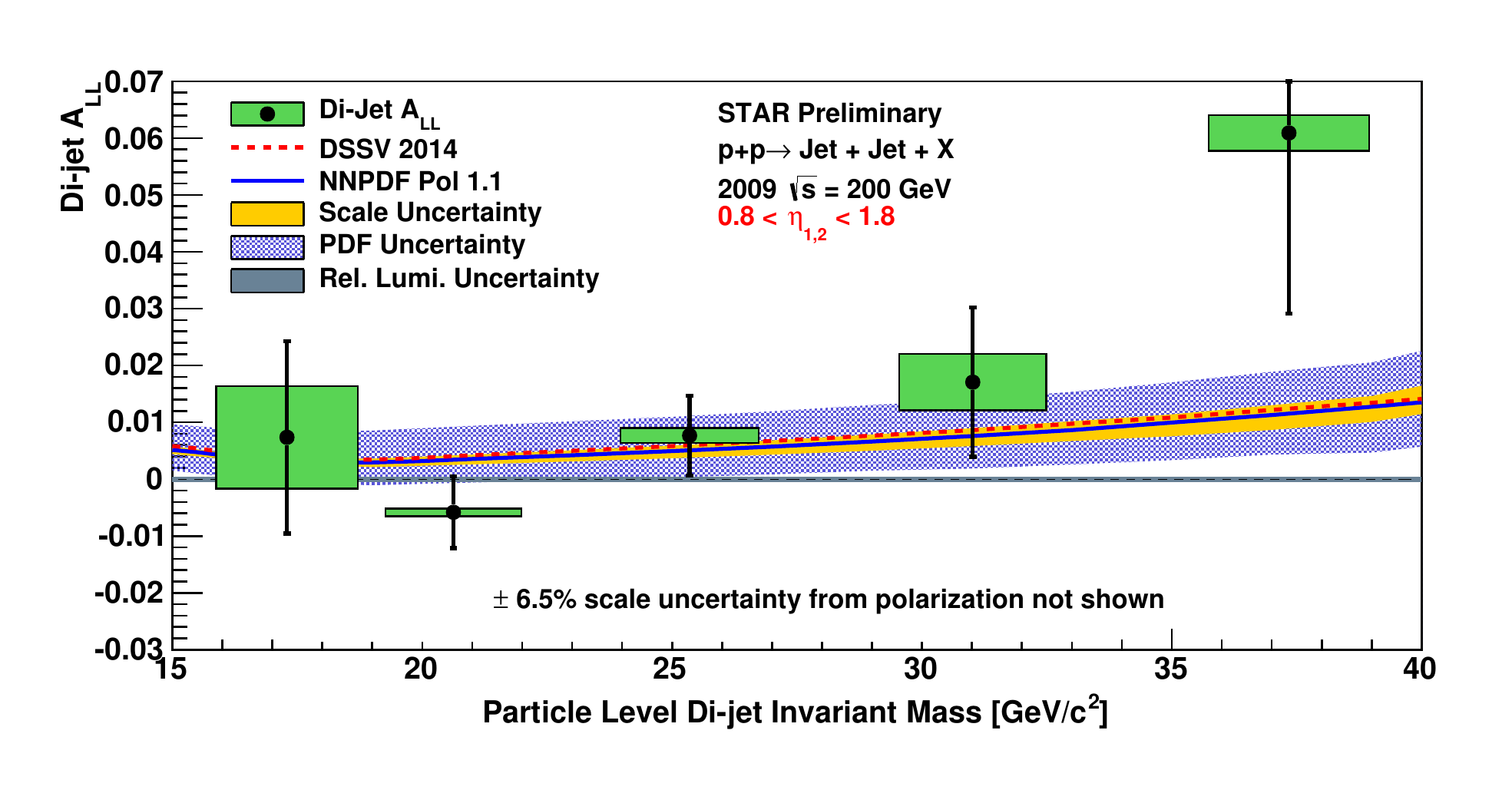}
    \caption{Preliminary Di-jet $A_{LL}$ for Endcap-Endcap \hskip 9em}
    \label{fig:ALL_EE}
  \end{minipage}
  \begin{minipage}{.495\textwidth}
    \centering
    \includegraphics[width=1.03\textwidth,height=1.2\textwidth]{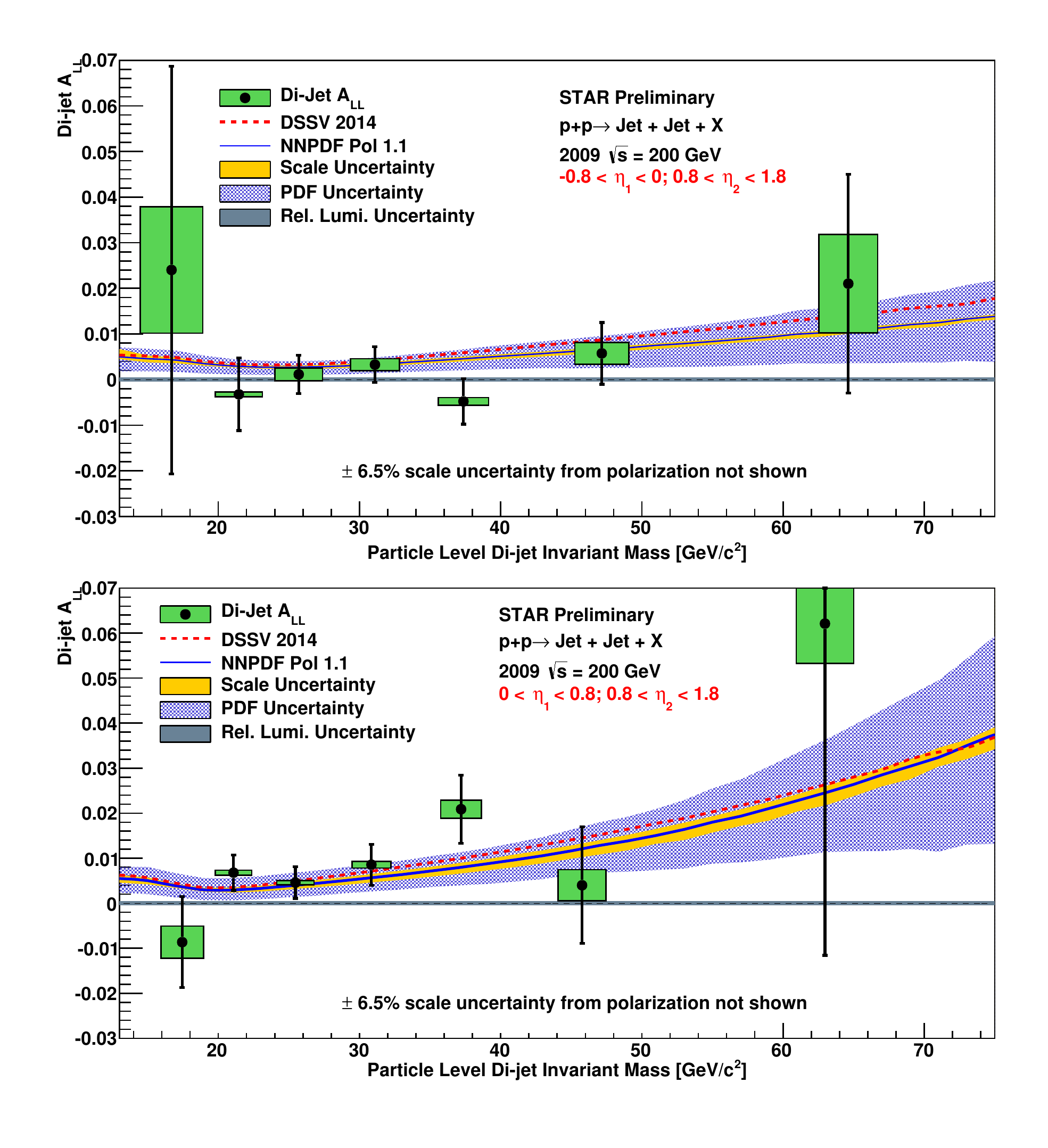}
    \caption{Preliminary Di-jet $A_{LL}$ for Barrel-Endcap different topologies}
    \label{fig:ALL_BE_DiffTopo}
  \end{minipage}
\end{figure*}

The Barrel-Endcap di-jet asymmetry $A_{LL}$ is presented for both full (Fig. \ref{fig:ALL_BE_FullTopo}) and distinct (Fig. \ref{fig:ALL_BE_DiffTopo}) topologies based on the range of the Barrel jet's pseudorapidity. A trigger and reconstruction bias correction was determined by comparing $A_{LL}$ from simulation at the detector and particle levels using several polarized PDFs \cite{STAR2009pp200Dijets}. The $A_{LL}$ shift was taken as the average values of the minimum and maximum $\Delta A_{LL} = A^{detector}_{LL} - A^{particle}_{LL}$ for the selected PDFs. Half of the difference between the minimum and maximum $\Delta A_{LL}$ was taken as a systematic error on the correction.\par
Figures \ref{fig:ALL_BE_FullTopo}, \ref{fig:ALL_EE} and \ref{fig:ALL_BE_DiffTopo} show the preliminary forward di-jet $A_{LL}$ values as a function of di-jet mass, which has been corrected back to the particle level. The heights of the uncertainty boxes represent the total systematic error due to trigger and reconstruction bias, combined with the error due to residual transverse polarization components in the beams. The relative luminosity uncertainty also results in a scaling uncertainty that is common to all points and is represented by the gray band on the horizontal axis. The widths of the uncertainty boxes represent the systematic error associated with the corrected di-jet mass values and, in addition to contributions from the uncertainty on the correction to the particle level, include the uncertainties on calorimeter tower gains and efficiencies, as well as TPC momentum resolution and tracking efficiencies. A further uncertainty was added in quadrature to account for the difference between the PYTHIA parton level and NLO CT10 \cite{nloct10} di-jet cross sections. Underlying events are also studied in both simulation and data, and are included in the total systematic errors.\par
Theoretical $A_{LL}$ values were obtained from the dijet production code of deFlorian \textit{et al.} \cite{TheoryCurve}, using the DSSV2014 \cite{PhysRevLett.113.012001} and NNPDFpol1.1 \cite{Nocera2014276} polarized PDF sets as input. Uncertainty bands representing the sensitivity to factorization and renormalization scale (solid) and polarized PDF uncertainty (hatched) were generated for the NNPDF results. Overall, the new data are seen to be in good agreement with current theoretical expectations. This suggests that incorporating these results into global analyses may not change the value of $\Delta g(x)$ significantly, but should lead to reduced uncertainties on this quantity, especially at lower Bjorken-$x$.\par
\section{Conclusion}
In summary, the preliminary forward region di-jet longitudinal double-spin asymmetries from STAR 2009 $pp$ collisions at $\sqrt s = 200\ \mathrm{GeV}$ are reported. The $A_{LL}$ results support the most recent DSSV and NNPDF predictions, which included 2009 RHIC inclusive jet and pion data. With the increased statistics from 2012 and 2013 at $\sqrt s = 510\ \mathrm{GeV}$, STAR data will help to better constrain the value and shape of $\Delta g(x)$ at lower Bjorken-$x$.\par
This work was supported in part by the US National Science Foundation.

\bibliography{spin2016_tinglin_Ref}
\end{document}